\documentclass[apj]{emulateapj}
\usepackage{apjfonts}

\newcommand\erg{{\rm erg}}
\newcommand\cm{{\rm cm}}
\newcommand\s{{\rm s}}
\newcommand\snr{G12.82${-}$0.02}
\newcommand\phndn{\phm{0.0}}

\shortauthors{Helfand, Becker, \& White}
\shorttitle{The TeV-Radio Connection}

\slugcomment{Submitted to Astrophysical Journal Letters}

\begin{document}

\title{A Radio Counterpart for the Unidentified TeV Source HESS J1813-178:\\
The Radio-Gamma-Ray Connection}

\author{David J. Helfand}
\affil{Columbia Astrophysics Laboratory, Columbia University,
550 West 120$^{th}$ Street, New York, NY 10027, USA}
\email{djh@astro.columbia.edu}

\author{Robert~H.~Becker}
\affil{Physics Department, University of California, 1 Sheilds Avenue, Davis, CA 95616;\\
and Institute of Geophysics and Planetary Physics, Lawrence Livermore National Laboratory,
Livermore, CA 94550}
\email{bob@igpp.ucllnl.org}

\and 

\author{Richard~L.~White}
\affil{Space Telescope Science Institute, 3700 San Martin Drive, Baltimore, MD 21218}
\email{rlw@stsci.edu}

\begin{abstract}

We discovered independently the shell-type supernova remnant \snr, 
recently reported by Brogan et al. (2005), which is coincident with the 
unidentified TeV gamma-ray source revealed in the HESS survey of the Galactic 
plane. Estimating the ambient starlight at the location of this source from the
integrated Ly$\alpha$ luminosity of the nearby \ion{H}{2} region W33, we conclude
that inverse Compton emission is a viable explanation for the observed TeV
emission. Examining remnants in the survey of Aharonian et al. (2005a) including
those detected above 200~Gev {\it and} those not detected, we find a strikingly 
large range of more than three orders of magnitude in the radio to TeV flux 
ratios. We briefly explore the possible explanations of this range and the 
implications for the TeV emission mechanism.

\end{abstract}


\section{Introduction}

The sites of cosmic ray acceleration in the Galaxy are widely believed to
be supernova remnants (SNRs), although the evidence for this belief has been
largely indirect. Over the last twenty years, nonthermal X-ray emission has been 
detected in several remnants, allowing us to infer that electrons with energies
up to at least 100 TeV are present (e.g., Cas A -- Allen et al.\ 1997; SN1006 -- 
Becker et al.\ 1980; Koyama et al.\ 1995; G$347.3-0.5$ -- Slane et al.\ 1999). Recently, the 
commissioning of the High Energy Stereoscopic System (HESS) for the detection of TeV 
gamma rays (Benbow et al.\ 2005) has opened a new window on high energy processes
in the Galaxy, providing a tool to locate directly hadronic CR acceleration
sites. A HESS survey of the innermost 60 degrees of Galactic longitude has 
recently been published (Aharonian et al.\ 2005a), and the results suggest we 
have much to learn about the details of CR acceleration in SNRs.

Of the ten discrete TeV-emitting sources found within the survey area
($-30^{\circ}<l<30^{\circ}, |b|<3^{\circ}$) only half have plausible 
identifications in existing source catalogs: the Galactic Center itself Sgr A* 
(Aharonian et al.\ 2004a), the X-ray synchrotron-dominated shell-type SNR
RX J1713.7-3946 (Muraishi et al.\ 2000; Aharonian 2004b), and three new sources
coincident with cataloged SNRs -- HESS J1640-465 with G338.3${-}$0.0, HESS J1804-216
with G8.7${-}$0.1, and HESS J1834-087 with G23.3${-}$0.3 -- although, given the inexact 
spatial coincidences and a probability of $>50\%$ for at least one chance 
HESS-SNR coincidence, not all of these identifications are secure.\footnote{The 
composite SNR G0.9${+}$0.1 (Helfand \& Becker 1987) was also detected by HESS
during a pointed observation of the Galactic Center (Aharonian et al.\ 2005b) but
it falls below the threshold of the HESS plane survey.} Even more interesting is
the fact that there are an additional 88 catalogued remnants (from Green 2004)
in the survey area,
none of which are seen. While the detected remnants G8.7${-}$0.1 and G23.3${-}$0.3 rank
third and fifth, respectively, in radio flux density at 1~GHz among known SNRs 
in the survey area, G338.3${-}$0.0 ranks only $42^{\it nd}$, and \snr\
(discussed below) would rank $86^{\it th}$.  Even given the possibility
of a false identification, it is clear that a bright radio flux
density is neither necessary nor sufficient to ensure detected TeV
gamma rays.

\section{A faint new SNR}

We are in the process of contructing a Multi-Array Galactic Plane Imaging
Survey (MAGPIS) with the Very Large Array\footnote{The VLA is operated by
the National Radio Astronomy Observatory which is a facility of the National 
Science Foundation operated under cooperative agreement by Associated
Universities, Inc.} operating at a wavelength of 20~cm (D.~J.~Helfand et al.,
2005, in preparation\footnote{The MAGPIS images are all publicly available on the
MAGPIS Web site at
\url{http://third.ucllnl.org/gps}.}). The images, constructed from B-, C-, and D-configuration 
observations plus single-dish data from the Effelsberg Galactic plane
survey (Reich, Reich \& Fuerst 1990) have a resolution of $\sim 
5^{\prime\prime}$, a sensitivity over most of the area of $\sim 1-2$~mJy, and a 
dynamic range of $\sim 1000:1$. The longitude range covered to date 
$(32^{\circ}>l>5^{\circ}$) includes five of the ten sources in the HESS Galactic
plane survey.

\begin{figure*}
\epsscale{0.80}
\plotone{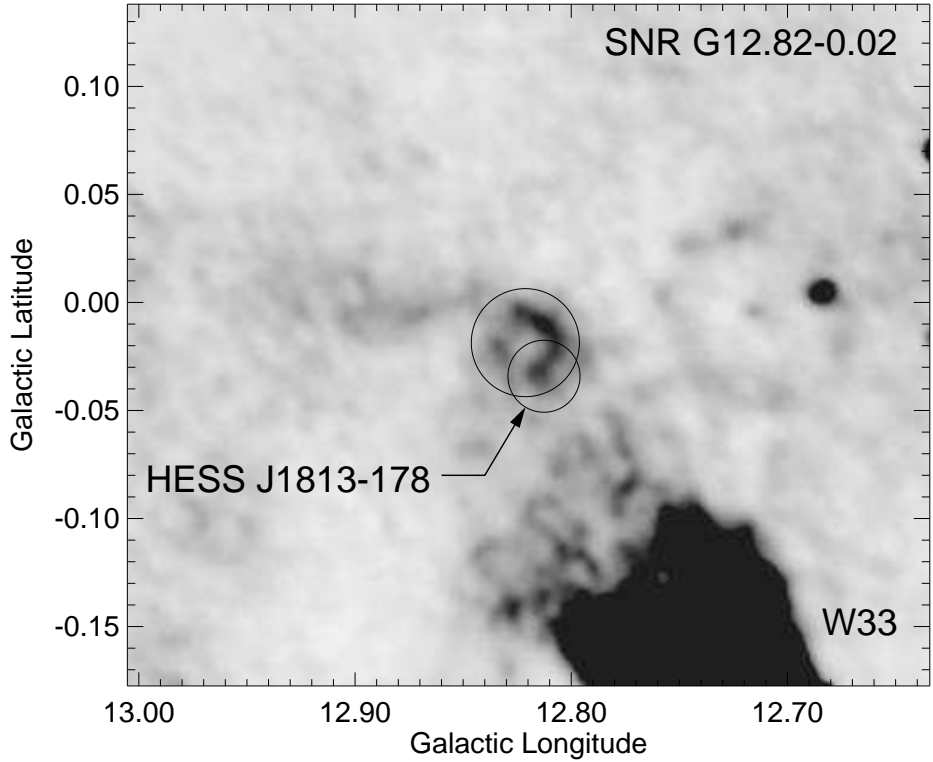}
\caption{
VLA 20~cm image of the TeV source HESS J1813-178. The $1^{\prime}$
radius TeV error circle is shown, and the supernova remnant \snr\
is marked by a $3^{\prime}$ diameter circle.  The bright nearby \ion{H}{2}
region W33 can be seen at the lower right.
}
\label{fig-radio}
\end{figure*}

Figure~\ref{fig-radio} shows a 20~cm greyscale plot of a $20^{\prime}$ region near
the TeV source HESS J1813-178. The TeV centroid is marked with a circle
whose radius is equal to the $1^{\prime}$ positional uncertainty in the
source centroid. The source is reported as extended, with a Gaussian
fit yielding $\sigma = 3^{\prime}$. We have drawn a circle with a diameter
of $3^{\prime}$ to highlight the circular shell of radio emission centered at
RA(J2000) = 18 13 35.5, Dec(J2000) = -17 49 40 (\snr); the center
of this shell is within $1^{\prime}$ of the TeV centroid.

This three-array-plus-single-dish image is a superset of the data analyzed
by Brogan et al.\ (2005) who recently reported the discovery of
this source. Within the uncertainties, our derived flux density value is
consistent with theirs: $S_{20} = 650$~mJy. Only ten of the 231 remnants in the 
most recent SNR catalog (Green 2004) are this faint. We also made a 90~cm map
from the archived data of Brogan et al.\ and independently completed an analysis 
of the {\it ASCA} image containing this source; again all our results are consistent
with theirs within the uncertainties, so we need not reiterate the results here.
We note, however, that we see no evidence for the X-ray ``soft excess'' claimed
by Ubertini et al.\ (2005) which is somewhat implausible given
the high column density to the source.

\section{Discussion}

\subsection{The environment of \snr}

\snr\ lies $\sim 8^{\prime}$ above (in Galactic latitude) the core
of the bright star-forming region W33. First studied in detail by
Haschick \& Ho (1983), the massive star cluster powering this complex
of \ion{H}{2} regions has a total Lyman continuum ionizing flux of $>10^{50}$
photons s$^{-1}$. From NH$_3$ observations, the authors infer gas
densities of $n({\rm H_2}) > 10^4$ cm$^{-3}$ and a total gas mass of
$\sim 10^5$ M$_{\odot}$. Several masers have been detected towards W33;
the dominant velocity is $\sim 36$~km s$^{-1}$ leading to a (near)
kinematic distance of $\sim 4$~kpc. Brogan et al.\ (2005) cite further
evidence that \snr\ is associated with this star formation
complex. At this distance, the source's intrinsic (unobscured) X-ray 
luminosity $L_x(0.1-10{\rm keV}) = 5 \times 10^{34}\,\erg\,\s^{-1}$, and the
gamma-ray luminosity $L_{\gamma}(>200~{\rm GeV}) = 1 \times 10^{34}\,\erg\,
\s^{-1}$. The radio luminosity is $L_r(10^7~{\rm to}~10^{11}~{\rm Hz}) = 3.6 
\times 10^{32}\,\erg\,\s^{-1}$ (Brogan et al.\ 2005).

\subsection{Gamma-ray emission mechanisms}

Several possible emission mechanisms can lead to TeV gamma-ray emission:
a) Inverse Compton scattering by off highly relativistic electrons of
ambient starlight, the Cosmic Microwave Background, and/or thermal dust photons, 
b) nonthermal bremsstrahlung of these same electrons, and c) pion production in
p-p collisions followed by $\Pi^0$ decay. The high-energy primaries are thought
to arise from diffusive shock acceleration in the outward moving blast wave from
the supernova event. Lower energy electrons produced in the same process power
the observed radio emission. Thus, it is not implausible that there should be
some relation between the radio and TeV fluxes of remnants.

\subsection{Inverse Compton Emission}

In discussing the TeV emission from the composite SNR G0.9${+}$0.1, Aharonian et al.\
(2005b) produce a model in which the X-rays arise as synchrotron emission by
relativistic electrons in the pulsar wind nebula, and the gamma rays arise from 
inverse Compton scattering of dust IR photons, the cosmic microwave background,
and ambient starlight off these same electrons. The starlight
dominates at the low energy end of the gamma-ray band, while the CMB
photons dominate at the high energy end (the dust emission provides less than
a 10\% contribution at all energies). The assumed starlight
energy density in the Galactic Center region is 5.7~eV cm$^{-3}$ and the
gamma-ray flux of the source is $f_{\gamma}(>200{\rm GeV}) =2.3 \times 10^{-12}
\,\erg\,\cm^{-2}\s^{-1}$ at 8.5~kpc; if moved to the 4~kpc distance of \snr,
the flux would be $10.4\times 10^{-12}\,\erg\,\cm^{-2}\s^{-1}$.

If we assume the starlight energy density in the vicinity of \snr\ is
dominated by the UV flux from W33, we have
\begin{equation}
\epsilon({\rm starlight}) = L E_{\nu} / 4 \pi d^2 c \quad ,
\end{equation}
and for $L = 10^{50}{\rm ph/s}, E_{\nu} = 15{\rm eV}, {\rm and}~d = 12$~pc
(the core of the \ion{H}{2} region is $12^{\prime}$ away), $\epsilon({\rm starlight})
\sim 3$~eV cm$^{-3}$ or roughly half the value for G0.9${+}$0.1. Ambient Galactic
starlight might increase this value by $\sim 25\%$, while extensive local
obscuration by dust will reduce it by an unknown amount. The gamma-ray
flux of $5 \times 10^{-12}\,\erg\,\cm^{-2}\s^{-1}$ is also roughly half
the value for G0.9${+}$0.1. Thus, within the rather large uncertainties
associated with several of these quantities, the synchrotron X-ray/IC gamma-ray
model appears to provide a consistent explanation for the high-energy emission
from \snr\ (cf. Figure 3 of Brogan et al.\ (2005) where the IC model using
CMB photons alone ($\sim 1$~eV cm$^{-3}$) falls roughly a factor
of three below the observations, consistent with the three-times higher photon
density implied by the proximity of W33). Since the CMB photons provide a
somewhat larger fractional contribution in the case of this source, one 
prediction of this explanation is that the gamma-ray spectrum of \snr\ 
should be somewhat flatter than the $\Gamma = 2.4$ found for G0.9${+}$0.1. The view 
of Brogan et al.\ (2005) that pion production from accelerated hadrons produces 
the TeV flux is also tenable.

\subsection{What makes TeV remnants?}

The discovery of the radio counterpart to HESS J1813-178, along with the
non-detection of several radio-bright remnants in the HESS Galactic plane
survey demonstrate a large spread in the radio-to-TeV flux ratios for
remnants. Adopting as a HESS survey threshold the photon fluxes of the
faintest sources detected, $S(>200 {\rm TeV})=9 \times 10^{-12}$ ph cm$^{-2}$
s$^{-1}$, and a spectral index of $\Gamma = -2.4$ yields a flux
threshold of $3.6 \times 10^{-12}\,\erg\,\cm^{-2}\s^{-1}$ at the Galactic
plane. We use the sensitivity curve in Figure 2 of Ahoranian et al.\ (2005a)
to scale up the limits with increasing Galactic latitude.

In Table~\ref{table-snr}, we provide data on several remnants including the detected TeV
sources and some of the brighter radio remnants which are not seen. We derive 
radio fluxes at 1~GHz from the Green (2004) catalog, assuming a uniform radio
spectral index of $\alpha \sim -0.65$ and integrating from $10^7~{\rm 
to}~10^{11}$~Hz. Changing the assumed value of $\alpha$ over the observed range 
for shell-type remnants from 0.4 to 0.8 translates to a flux density range of 
a factor of 2.5. Gamma-ray fluxes ($E>200$~GeV) are taken from the sources 
cited; upper limits for sources undetected in the HESS plane survey are set 
at the flux of faintest detected objects. The final column of the table 
lists the ratio of radio to gamma-ray flux. There is an extraordinary 
range of nearly a factor of $10^4$ in this value.

There is no obvious correlation of the TeV-radio flux ratio with angular or
physical remnant size or with radio spectral index; e.g., G6.4${-}$0.1 and G8.7${-}$0.1
are within a factor of two of the same diameter and have similar spectral
indices of $\alpha \sim0.5$ and yet differ in their flux ratios by a factor of
16. What parameters do the TeV-emitting sources have in common?

Of the six detected remnants in the HESS survey area, two (G347.3${-}$0.5 and 
\snr) have X-ray spectra dominated by a power law component assumed 
to arise in synchrotron emission associated with the SNR shell,
while a third (G0.9${+}$0.1) has a central pulsar
wind nebular that also emits synchrotron X-rays; the other three remnants have not
been studied at X-ray wavelengths. Of the remnants undetected by HESS listed in 
Table~\ref{table-snr}, thermal X-ray emssion dominates in G6.4${-}$0.1,
G21.8${-}$0.6 was marginally detected by {\it Einstein}, and has not
been observed since, and G348.5${+}$0.1 is 
undetected in X-rays. However, X-rays have been detected from some two dozen other 
remnants in the survey area and eighteen are dominated by soft, thermal
emission. There is a clear preference for TeV emitters to also show substantial
synchrotron X-ray emission implying the presence of electrons with
energies significantly in excess of 100 TeV; likewise, strong synchrotron X-ray 
emission is lacking in many of the undetected remnants.

\begin{deluxetable*}{lccccc}
\tablecolumns{6}
\tablewidth{0pt}
\tabcolsep=3pt
\tablecaption{Radio and Gamma-ray Properties of Selected Remnants}
\tablehead{
\colhead{SNR} & \colhead{Size} & \colhead{Distance} &
\colhead{$f_r$} & \colhead{$f_{\gamma}$} & \colhead{$f_{\gamma}/f_r$} \\
 & \colhead{(arcmin)} & \colhead{(kpc)} &
\multicolumn{2}{c}{($10^{-12}\,\erg\,\cm^{-2}\s^{-1}$)} & }
\startdata
G0.9${+}$0.1       & \phn8 & \phndn8.5 &   2.5 &  2.3    & 0.9    \\
G6.4${-}$0.1 (W28) &    48 & \phn2     &   43  &  $<3.6$ & $<0.1$ \\
G8.7${-}$0.1 (W30) &    26 & \phn6     &   11  &  16     & 1      \\
G12.82${-}$0.02    & \phn3 & \phn4     &   0.2 &  12     & 60    \\
G21.8${-}$0.6      &    20 & $>6$\phn  &   9.5 &  $<3.7$ & $<0.4$ \\
G23.3${-}$0.3 (W41)&    38 & \phn5     &   9.6 &  13     & 1      \\
G338.3${-}$0.0     & \phn8 & \nodata   &   1.0 &  19     & 20     \\
G347.3${-}$0.5     &    60 & \phn6     & $>0.5$ & 250    & $<500$ \\
G348.5${+}$0.1     &    15 & 10        &   10  &  $<3.6$ & $<0.4$ \\
\enddata
\tablecomments{This table includes SNRs in the HESS survey that are
detected as TeV sources and/or are radio bright. The fluxes are for
0.2--10 TeV ($f_\gamma$) and $10^7$--$10^{11}$~Hz ($f_r$).}
\label{table-snr}
\end{deluxetable*}

% For G0.9, PWN may be TeV sources (IC model works)
% For G8.7 -- I don't see SNR!
% For G23.3 -- is coincident with inner shell may not be whole f_r
% For G347 -- only have flux density of brightest filaments ~10% of remnant

\begin{deluxetable*}{lcccc}
\tablecolumns{5}
\tablewidth{0pt}
\tabcolsep=3pt
\tablecaption{X-ray, Gamma-ray, and Radio Properties for Pulsar Wind Nebulae in the HESS Survey}
\tablehead{
\colhead{SNR} & \colhead{Gamma-ray Flux} & \colhead{X-ray Flux} &
\colhead{Radio Flux Density} & \colhead{Flux in Starlight} \\
& \multicolumn{2}{c}{($10^{-12}\,\erg\,\cm^{-2}\s^{-1}$)} &
(Jy) & \colhead{($10^{-12}\,\erg\,\cm^{-2}\s^{-1}$)}}
\startdata
G0.9${+}$0.1    & \phn\phn2.3  &   22      &  4.5\phn & \phn\phn2.1    \\
G11.2${-}$0.3   & $<3.7$       &   10      &  0.36    & $<7.4$         \\
G16.7${-}$0.3   & $<3.7$       &   \phn2   &  0.12    & $<37$          \\
G20.0${-}$0.2   & $<3.6$       &   \phn1   &  9.7\phn & $<72$          \\
G21.5${-}$0.9   & $<4.5$       &  110\phn  &  6.0\phn & $<0.8$         \\
G29.7${-}$0.3   & $<3.9$       &   40      &  0.28    & $<2.0$         \\
G343.1${-}$2.3  & $<9.0$       &    0.005  &  0.03    & $<3\times10^4$ \\
G12.82${-}$0.02\tablenotemark{a} & \phn\phn5.2 & 115\phn & 0.65 & \phn\phn0.9 \\
\enddata
\tablecomments{This table includes SNRs in the HESS survey that are PWN.
Gamma-ray fluxes are for the 0.2--10 TeV band.
X-ray fluxes are for the 0.5--10 keV band.
Radio flux densities are at $\lambda=20$~cm.
The X-ray and radio values for composite remnants include only the PWN.
Starlight fluxes are $\nu f_\nu$ at 15~eV determined from Eqn.~(2).
\tablenotetext{a}{Not known to be a PWN, but included for comparison.}
}
\label{table-pwn}
\end{deluxetable*}

Another set of SNRs that are known to have synchrotron X-ray spectra
are the pulsar wind nebulae (PWNe) and composite objects that contain
a PWN.  The parameters of PWNe falling in the HESS survey are listed
in Table~\ref{table-pwn} (including the HESS-detected SNR G0.9${+}$0.1,
which is repeated from Table~\ref{table-snr}.) The gamma-ray upper
limits in the $0.2-10$~TeV band are derived as described above. The
X-ray data are taken from the literature and scaled to be unabsorbed
fluxes in the $0.5-10$~keV band. The radio flux densities are for
20~cm; in composite remnants, the X-ray and radio values are for
the PWN only. \snr\ is included as the last entry in the table for
comparison.

If gamma rays were in general produced by hadronic processes, one
might naively assume there would be some relationship between the
radio flux density of a remnant which ultimately arises from the
particle acceleration process, and the gamma-ray flux. Howver, as
noted above, the brightest radio remnants remain undetected while
some of the faintest remnants are seen. For example, in Table~\ref{table-pwn},
the radio flux densities and gamma-ray fluxes for G0.9${+}$0.1 and 
\snr\ seem unrelated; G0.9${+}$0.1 is seven times brighter in the 
radio and only half as bright in the TeV range. However, examining
the X-ray and gamma-ray fluxes in the context of an inverse Compton
model for gamma-ray emission appears somewhat more promising.

The final column of Table~\ref{table-pwn} lists limits on the flux density 
$\nu f_{\nu}$ in starlight 
at the PWN under the assumption that the IC model presented in
Aharonian et al.\ (2005b) for G0.9${+}$0.1 is correct, and the highly
simplified additional assumption that the gamma-ray-to-x-ray flux
ratio for these sources is universal except that it scales with the
ambient starlight flux; i.e.,
\begin{equation}
f_{\gamma}/f_x = 0.05 \, f_{opt}\quad,
\end{equation}
where the constant is derived from the parameters of G0.9${+}$0.1. It
is apparent that these assumptions are not seriously
challenged by the available data. As noted above, the {\it observed}
value of the ambient flux from the \ion{H}{2} region W33 is sufficient to
explain the gamma-rays from \snr\ with this relation. Furthermore,
all the other PWN have upper limits on their starlight fluxes that are plausible;
most are far above the high value of $2 \times 10^{-12}\,\erg\,\cm^{-2}\s^{-1}$
found near the Galactic Center. Only G21.5${-}$0.9 is significantly
lower; it is even slightly below the value for \snr, but an examination 
of the region surrounding this remnant shows no bright \ion{H}{2} regions lie within
$10^{\prime}$ and, given the source's location nearly
one degree off the Galactic plane, this lower starlight flux is expected.

\section{Conclusions}

One of the blank-field TeV-emitting gamma-ray sources in the HESS Galactic 
plane survey is firmly identified with a small-diameter, faint, shell-like
radio SNR which has an X-ray spectrum dominated by synchrotron X-rays. The
gamma-ray flux can be explained by inverse Compton scattering of the
starlight from the nearby \ion{H}{2} region W33 off the X-ray emitting electrons.
The detections and upper limits for the 90 other supernova remnants in
the inner Galaxy demonstrate that there is a large range ($>10^3$) of 
radio-to-gamma-ray flux ratios, and an association of detectable TeV 
emission with X-ray synchrotron emission. Examination of the seven pulsar
wind nebulae in the survey region demonstrates consistency with the notion
that the range of gamma-ray-to-x-ray flux ratios can be explained by
variations in the energy density in ambient starlight. Deeper HESS 
observations, as well as sensitive X-ray observations of the many remnants
whose X-ray emission remains uncharacterized will be necessary before
more informed contraints on the TeV emission mechanism can be constructed.

\acknowledgments

D.J.H.\ and R.H.B.\ acknowledge the support of the National Science Foundation
under grants AST~02-6309 and AST~02-655, respectively.  D.J.H.\ was
also supported in this work by NASA grant NAG5-13062.  R.H.B.'s work
was supported in part under the auspices of the US Department of
Energy by Lawrence Livermore National Laboratory under contract
W-7405-ENG-48.  R.L.W.\ acknowledges the support of the Space Telescope
Science Institute, which is operated by the Association of Universities
for Research in Astronomy, Inc., under NASA contract NAS5-26555.


\begin{thebibliography}{}

\bibitem[Aharonian et al.(2004a)]{2004A&A...425L..13A} Aharonian, F., et 
al.\ 2004a, \aap, 425, L13 
 
\bibitem[Aharonian et al.(2004b)]{2004Natur.432...75A} Aharonian, F.~A., et 
al.\ 2004b, \nat, 432, 75 

\bibitem[Aharonian et al.(2005a)]{2005Sci...307.1938A} Aharonian, F., et 
al.\ 2005a, Science, 307, 1938 

\bibitem[Aharonian et al.(2005b)]{2005A&A...432L..25A} Aharonian, F., et 
al.\ 2005b, \aap, 432, L25 

\bibitem[Allen et al.(1997)]{1997ApJ...487L..97A} Allen, G.~E., et al.\ 
1997, \apjl, 487, L97 

\bibitem[Becker et al.(1980)]{1980ApJ...240L..33B} Becker, R.~H., 
Szymkowiak, A.~E., Boldt, E.~A., Holt, S.~S., \& Serlemitsos, P.~J.\ 1980, 
\apjl, 240, L33 

\bibitem[Benbow \& Hess Collaboration(2005)]{2005AIPC..745..611B} Benbow, 
W., \& Hess Collaboration 2005, AIP Conf.~Proc.~745: High Energy Gamma-Ray 
Astronomy, 745, 611 

\bibitem[Brogan et al.(2005)]{2005astro.ph..5145B} Brogan, C.~L., Gaensler, 
B.~M., Gelfand, J.~D., Lazendic, J.~S., Lazio, T.~J., Kassim, N.~E., \& 
McClure-Griffiths, N.~M.\ 2005, ArXiv Astrophysics e-prints, 
arXiv:astro-ph/0505145 

\bibitem[Green(2004)]{2004BASI...32..335G} Green, D.~A.\ 2004, Bulletin of 
the Astronomical Society of India, 32, 335 

\bibitem[Haschick \& Ho(1983)]{1983ApJ...267..638H} Haschick, A.~D., \& Ho, 
P.~T.~P.\ 1983, \apj, 267, 638 

\bibitem[Helfand \& Becker(1987)]{1987ApJ...314..203H} Helfand, D.~J., \& 
Becker, R.~H.\ 1987, \apj, 314, 203 

\bibitem[Koyama et al.(1995)]{1995Natur.378..255K} Koyama, K., Petre, R., 
Gotthelf, E.~V., Hwang, U., Matsuura, M., Ozaki, M., \& Holt, S.~S.\ 1995, 
\nat, 378, 255 

\bibitem[Muraishi et al.(2000)]{2000A&A...354L..57M} Muraishi, H., et al.\ 
2000, \aap, 354, L57 

\bibitem[Reich et al.(1990)]{1990A&AS...83..539R} Reich, W., Reich, P., \& 
Fuerst, E.\ 1990, \aaps, 83, 539 

\bibitem[Slane et al.(1999)]{1999ApJ...525..357S} Slane, P., Gaensler, 
B.~M., Dame, T.~M., Hughes, J.~P., Plucinsky, P.~P., \& Green, A.\ 1999, 
\apj, 525, 357 

\end{thebibliography}
\end{document}